\documentclass{jfm}

\usepackage{lipsum}                     
\usepackage{xargs}                      
\usepackage[dvipsnames]{xcolor}  
\usepackage{graphicx}
\usepackage{epstopdf, epsfig}
\usepackage{subcaption}
\usepackage{url}
\usepackage{lscape}
\usepackage{rotating}
\usepackage{soul}
\usepackage[english]{babel}                     
\usepackage{blindtext}                          
\usepackage{color}                              
\usepackage{comment}                            
\usepackage{wasysym}                            
\title{Extreme-scale motions in turbulent plane Couette flows}
\shorttitle{Extreme-scale motions in turbulent plane Couette flows}

\shortauthor{M. Lee and R. D. Moser}
\author{Myoungkyu Lee\aff{1}
 \and Robert D. Moser\aff{1,2}\corresp{\email{rmoser@ices.utexas.edu}}}

\affiliation{\aff{1}Center for Predictive Engineering and Computational
  Sciences, Insititute for Computational Engineering and Sciences, The
  University of Texas at Austin, TX 78712, USA
  \aff{2}Department of Mechanical Engineering, The University of Texas
  at Austin, TX 78712, USA}

\begin{document}

\maketitle

\begin{abstract}
We study the large-scale motions in turbulent plane Couette flows at
moderate friction Reynolds number up to $Re_\tau=500$.  Direct
numerical simulation domains were as large as $100\pi\delta \times
2\delta \times 5\pi\delta$, where $\delta$ is half the distance
between the walls. The results indicate that there are streamwise
vortices filling the space between the walls that remain
correlated over distances in the streamwise direction that increase
strongly with the Reynolds number, so that for the largest Reynolds
number studied here, they are correlated across the entire
$100\pi\delta$ length of the domain. The presence of these very long
structures is apparent in the spectra of all three velocity components
and the Reynolds stress.  In DNS using a smaller domain, the large
structures are constrained, eliminating the streamwise variations
present in the larger domain. Near the center of the domain, these
large scale structures contribute as much as half of the Reynolds
shear stress.
\end{abstract}


\section{Introduction}
\label{sec:intro}
Planar Couette and Poiseuille flows are the simplest canonical
configurations for the computational study of wall-bounded
turbulence. Since \citet{Kim:1987ub} carried out direct numerical
simulation (DNS) of Poiseuille flows at $Re_\tau = 180$, DNS has been
widely used to study the physics of wall-bounded turbulence at
Reynolds numbers ($Re$) up to
$Re_\tau=5200$ \citep{Moser:1999ty,delAlamo:2004bd,Hoyas:2006kl,LozanoDuran:2014kr,Bernardini:2014kh,Lee:2015er}.
Even higher $Re$ flows have been studied experimentally in pipes at
$Re_\tau = 10^5$ \citep{Hultmark:2012ce} and zero-pressure-gradient
(ZPG) boundary layers at $Re_\tau = 6.5\times
10^5$ \citep{Marusic:2010bb}.

On the other hand, the study of planar Couette flow has been more
limited in number of studies and Reynolds number, both experimentally
and computationally.  One of the reasons is the existence of very
large-scale motions in Couette flow that are particularly challenging
to represent in DNS. The first DNS study of turbulent Couette flows
was at $Re_\tau = 170$ by \citet{Lee:1991wt}. They used a periodic
computational domain size of $4\pi\delta \times 8/3\pi\delta$ in the
streamwise and spanwise directions, respectively, where $\delta$ is
half the distance between the walls. In their simulation, the most
energetic motion at the center of the domain is at wavenumbers
$k_x\delta=0$ and $k_z\delta = 1.5$ which has no variation in the
streamwise direction and is clearly an artifact of the finite
simulation domain size. \citet{Komminaho:1996fu} also performed DNS at
$Re_\tau = 52$ with a simulation domain of $28\pi\delta \times
8\pi\delta$ and obtained energy spectra with DNS peaks at the same
wavenumbers. Later, \citet{Tsukahara:2006bg,Avsarkisov:2014go,Pirozzoli:2014bm}
performed DNS at Reynolds numbers up to $Re_\tau=126$
($64\delta \times 6\delta$), $Re_\tau = 550$ ($20\pi\delta \times
6\pi\delta$) and $Re_\tau = 986$ ($18\pi\delta \times 8\pi\delta$),
respectively. Their results show that the wavelength of large-scale
motion is approximately $\lambda_z = 5\delta$. However, the domain
sizes for the previous works were not long enough to determine the
extent of the large-scale motion. On the other hand, the experimental
study with moving belts by \citet{Kitoh:2008gj,Kitoh:2005tn} at
$Re_\tau = 192$ shows that the extent of the large-scale motion in the
streamwise direction is $\lambda_x \approx 40\delta - 60\delta$.

Large-scale motions in transition in planar Couette flows have also
been observed. For example, \citet{Tillmark:1992kx} performed moving
belt experiments with an aspect ratio $L_x/\delta=150$ and observed
growing turbulent spots with the streamwise extent of about
$60\delta$. Using a moving belt facility with aspect ration
385, \citet{Prigent:2003gg} observed spots with streamwise coherence
over approximately $250 \delta$. More recently, using particle image
velocimetry (PIV) in an aspect ration 100 belt
experiment, \citet{Couliou:2015fw} studied growth of large-scale
motions.  Finally, using DNS, \citet{Duguet:2010dv} studied the
formation of turbulent spots with $800\delta$ long simulation domains.

The goal of the current study is to study the very large-scale motions
in Couette flow by simulating them in computational domains large
enough to allow their streamwise extent to be
determined. Through the discussion, similarity to and
contrasts with turbulent Poiseuille flow will also be made through
reference to the DNS study by \citet{Lee:2015er} and \citet{LozanoDuran:2014kr}. When
comparing dissimilar flows, there is always ambiguity as to how to
consistently compare Reynolds numbers. Here the Reynolds number
$Re_\tau$ is defined based on friction velocity and the domain
half-width $\delta$, as is common in Poiseuille flow. However, in a
study of transitional flows, \citet{Barkley:2007cn} suggest that the appropriate
length scale in Couette flow is the full width, since the shear does
not change sign across the layer. Assessing appropriate length
scales for comparison to Poiseuille flow would require Couette flow
simulations at much higher Reynolds number than reported here, which is
out of scope for the current study. For our purposes, it is sufficient
to note that the Couette flows studied here could be comparable to
Poiseuille flows with Reynolds number equal to or up to twice as high
as the Couette Reynolds number. In the remainder of this paper, we
present details of the current simulations (\S\ref{sec:method})
followed by simulation results (\S\ref{sec:result}). Finally, we
provide concluding remarks in \S\ref{sec:conclusion}.

\section{Numerical Simulations}
\label{sec:method}
In the discussion to follow, $x$, $y$ and $z$ denote the streamwise,
wall-normal and spanwise directions, respectively, and the
corresponding velocity components are denoted as $u$, $v$, and $w$,
respectively. Angle brackets $\langle \psi \rangle$ indicate the
expected value or average of the quantity $\psi$ over $x$ and $z$
directions and time, $t$. In addition, quantities with prime denote
the fluctuation from the average, i.e. $\psi' = \psi
- \langle \psi \rangle$ and $\langle \psi' \rangle = 0$.

We performed direct numerical simulations of incompressible turbulent
Couette flow by solving the Navier-Stokes equations using the
velocity-vorticity formulation introduced by
\citet{Kim:1987ub}. The flow is driven by two parallel planes moving in
opposite directions at constant speed, and there is no mean pressure
gradient. We specify a finite rectangular simulation domain of size
$L_x\times L_z$ with periodic boundary conditions in the wall-parallel
($x$ and $z$) directions. Boundary conditions at the walls are no-slip
and no-penetration. A Fourier-Galerkin discretization with $N_x\times
N_z$ Fourier modes is used in the wall-parallel directions, with
effective resolution defined by the Nyquist grid spacing $\Delta
x=L_x/N_x$ and $\Delta z=L_z/N_z$.  In the wall-normal ($y$)
direction, we use a seventh order basis spline (B-spline) collocation
method, with knots defined by:
\begin{equation}
\displaystyle
y_{i,\mathrm{knot}} = \delta \sin \left(\frac{i\pi}{N_y-7}-\frac{\pi}{2}\right), \quad i = 0, 1, 2, \cdots (N_y-7)
\label{eq:grid_stretch}
\end{equation}
where $N_y$ is the number of B-splines in the representation and the
number of collocation points.  Note that the knots near the walls are
denser than knots at the center of domain. The collocation points are
defined as the Greville
abscissae \citep{Johnson:2005ge}. See \citet{Lee:2013kv,Lee:2014ta,Lee:2015er}
for more information about the numerical method.

\begin{table}
\begin{center}
\def~{\hphantom{0}}
\resizebox{\textwidth}{!}{
\begin{tabular}{lcccccccccccc}
Case & $Re_\tau$ & $L_x/\delta$ & $L_z/\delta$ & $N_x$ & $N_y$ & $N_z$ &
$\Delta x^+$ & $\Delta y^+_w$ & $\Delta y^+_c$ &
$\Delta z^+$ & $Tu_{\tau}/\delta$ & Reference \\ [3pt]
\hline \hline
R93S  & ~92.93 & ~20$\pi$ & 5$\pi$ & ~~640 & 128 & ~288 & ~9.12 & 0.031 & 2.41 & 5.07 & 197.94 & Current\\
R93L  & ~92.91 & 100$\pi$ & 5$\pi$ & ~3072 & 128 & ~288 & ~9.50 & 0.031 & 2.41 & 5.06 & 185.83 & Current\\
R220S & 219.15 & ~20$\pi$ & 5$\pi$ & ~1280 & 192 & ~768 & 10.76 & 0.032 & 3.72 & 4.48 & 187.16 & Current\\
R220L & 219.48 & 100$\pi$ & 5$\pi$ & ~6144 & 192 & ~768 & 11.22 & 0.032 & 3.73 & 4.49 & 164.61 & Current\\
R500S & 502.63 & ~20$\pi$ & 5$\pi$ & ~3072 & 256 & 1536 & 10.28 & 0.040 & 6.34 & 5.14 & 150.79 & Current\\
R500L & 501.37 & 100$\pi$ & 5$\pi$ & 15360 & 256 & 1536 & 10.25 & 0.040 & 6.33 & 5.13 & 150.79 & Current\\
\hline
R125A & 131.60 & ~20$\pi$ & 6$\pi$ & ~~576 & 151 & ~384 & 14.36 & 0.746 & 2.14 & 6.46 & ~27.1~ & \citet{Avsarkisov:2014go} \\
R180A & 177.29 & ~20$\pi$ & 6$\pi$ & ~~864 & 251 & ~512 & 12.89 & 0.297 & 1.88 & 6.53 & ~34.~~ & \citet{Avsarkisov:2014go} \\
R250A & 243.74 & ~20$\pi$ & 6$\pi$ & ~1536 & 251 & ~768 & ~9.97 & 0.408 & 2.59 & 5.98 & ~53.~~ & \citet{Avsarkisov:2014go} \\
R550A & 553.43 & ~20$\pi$ & 6$\pi$ & ~2592 & 251 & 1536 & 13.42 & 0.926 & 5.88 & 6.79 & ~57.~~ & \citet{Avsarkisov:2014go} \\
\hline
R550P & 546.75 & ~60$\pi$ & 6$\pi$ & ~8192 & 257 & 2048 & 12.58 & 0.041 & 6.71 & 5.03 & ~~5.~~ & \citet{LozanoDuran:2014kr} \\
\end{tabular} }
\caption{Summary of simulation parameters. $N_x$ and $N_z$ are the number of Fourier modes, and $N_y$ is the number of collocation points. $\Delta y_w$ and $\Delta y_c$ of current work are the knot spacings at the wall and center line, respectively whereas $\Delta y_w$ and $\Delta y_c$ of works by \citet{Avsarkisov:2014go} and \citet{LozanoDuran:2014kr} are the collocation point spacing. $Re_\tau =
u_\tau \delta/\nu$ is the friction Reynolds number, and $Tu_\tau/\delta$
is the scaled total averaging time.}
\label{table:simulation_parameter}
\end{center}
\end{table}

We simulated six cases with three different Reynolds numbers,
$Re_\delta = U_w \delta / \nu$ where $U_w$ and $\nu$ are wall-speed
and kinematic viscosity, respectively. The domain size in the
streamwise direction was either $L_x=100\pi\delta$ or $20\pi\delta$,
while the domain size in the spanwise direction was $L_z=5\pi\delta$
for all cases. We use $L_x = 20\pi\delta$ for the small domain cases
because this is the largest streamwise length used in previous DNS
studies \citep{Avsarkisov:2014go}. In the spanwise direction, $L_z =
5\pi\delta$ is smaller than the domain size used
by \citet{Pirozzoli:2014bm} ($L_z = 8\pi\delta$), but it has been
found to be sufficient to capture the large scale motions in the
spanwise directions, as will be discussed in the next section. The
resolution ($\Delta x^+$, $\Delta y^+$ and $\Delta z^+$) was chosen to
be sufficient to represent the smallest dissipative motions, based on
experience in other wall-bounded turbulent flows \citep[see {\it
e.g.}][]{Lee:2015er}.  Table~\ref{table:simulation_parameter} catalogs
the parameters defining the cases discussed here.  The friction
Reynolds number, $Re_\tau$, is based on the friction velocity,
$u_\tau=\sqrt{\tau_w/\rho}$, where $\tau_w$ is the mean shear stress
at the wall and $\rho$ is the fluid density.  As is common in
wall-bounded turbulence, a superscript $+$ designates a quantity
normalized in wall units (i.e. normalized by $u_\tau$ and $\nu$).

Conservation of mean momentum requires that in stationary turbulent
Couette flow, the total stress, which is the sum of the Reynolds and
viscous shear stresses is a constant independent of $y$; particularly:
\begin{equation}
\tau_\mathrm{total}^+ = \left\langle\frac{\partial u
}{\partial y}\right\rangle^+ - \langle u'v' \rangle^+ = 1
\label{eq:total}
\end{equation}
To ensure that the simulation results were statistically stationary,
the total stress was monitored for consistency with (\ref{eq:total}).
In all the simulation results presented here, the deviation from this
analytic result was less than 0.2\%, and less than the estimated
sampling uncertainty determined by the technique outlined
in \citet{Oliver:2014dh}.  Estimated uncertainties in one-point
statistics are included in Appendix~\ref{sec:appendix}.  In the
following, we include data from \citet{Avsarkisov:2014go} for
comparison, when possible.

\section{Results}
\label{sec:result}
\begin{figure}
  \centering
  \includegraphics[width=\textwidth]{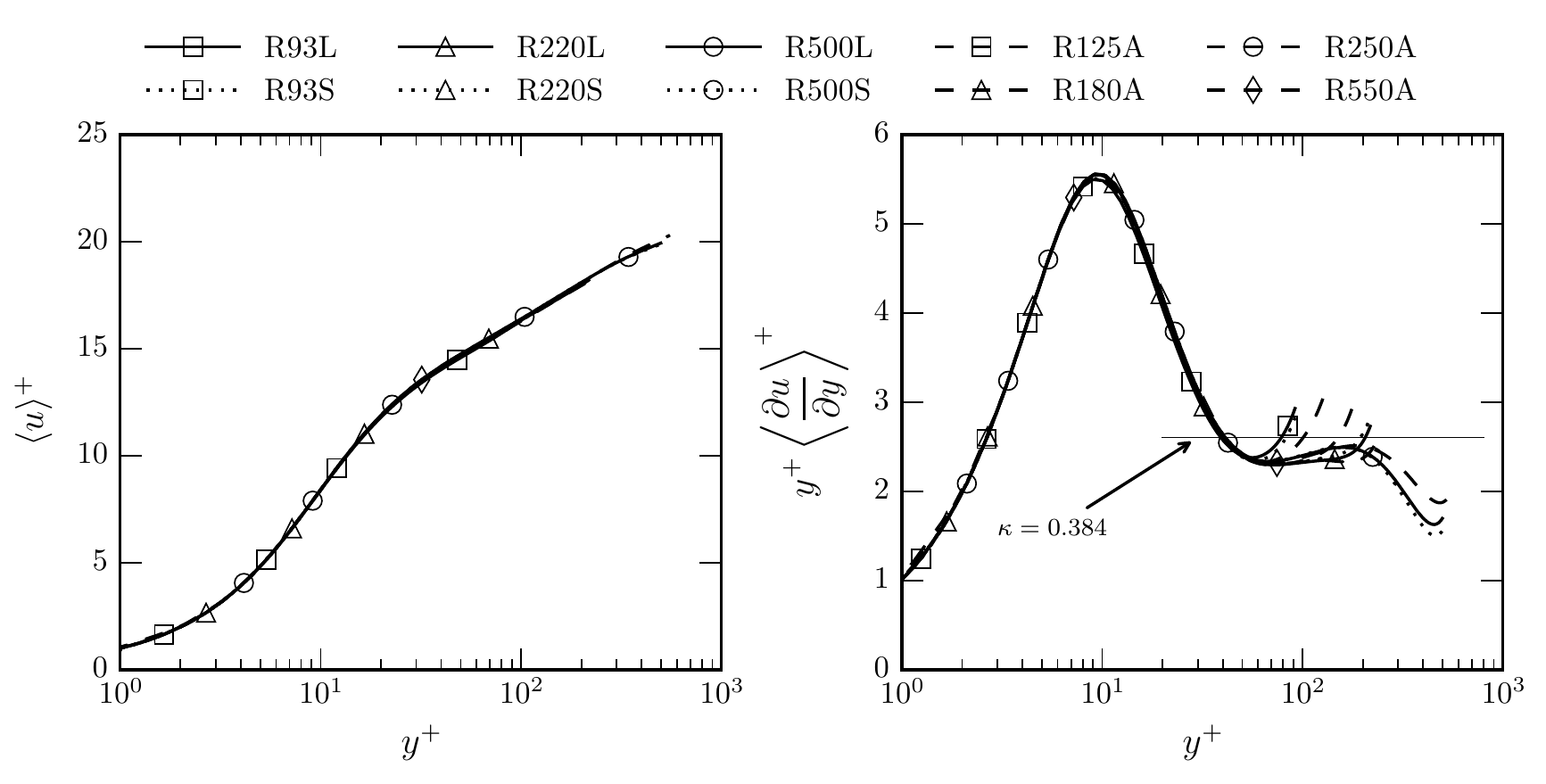}
  \caption{Mean velocity profile (left) and log-layer diagnostic function (right); Cases are defined in table~\ref{table:simulation_parameter}.}
  \label{fig:profile}
\end{figure}

The mean velocity profiles for each Couette flow case are shown in
figure~\ref{fig:profile}. Both the mean velocity and the log-layer
diagnostic $\beta=y^+ \langle \partial u^+/\partial y^+ \rangle$ are
shown. As expected, the Reynolds numbers are too low to produce a
convincing log-layer (constant $\beta$). Indeed in Poiseuille flow, an
unambiguous log layer does not appear until $Re_\tau\approx 5000$
\citep{Lee:2015er}.  Nonetheless, $\beta$ is more sensitive to small
variations in the mean profile, and therefore it is a good diagnostic
for detecting such differences. The $\beta$ profiles are consistent
among all the cases up to $y^+ \approx 50$. Beyond this, $\beta$
develops a plateau as $Re$ increases. However, at $Re_\tau\approx500$,
$\beta$ increases slowly until $y/\delta \approx 0.3$ then decreases
sharply, as previously observed by
\citep{Avsarkisov:2014go,Orlandi:2015bl}. At the low Reynolds numbers,
there is no discernible difference in $\beta$ between the small and
large domain cases, but there is a non-negligible difference away from
the wall in the R500 cases. In all cases, $\beta$ rolls up at the
center, even in the higher Reynolds number cases in which it drops
sharply above $y/\delta\approx 0.3$. This is required since at the
center $d\beta/dy=\langle\partial u/\partial y\rangle>0$ because
$\langle\partial^2 u/\partial y^2 \rangle=0$ by symmetry.

\begin{figure}
  \centering
  \includegraphics[width=.5\textwidth]{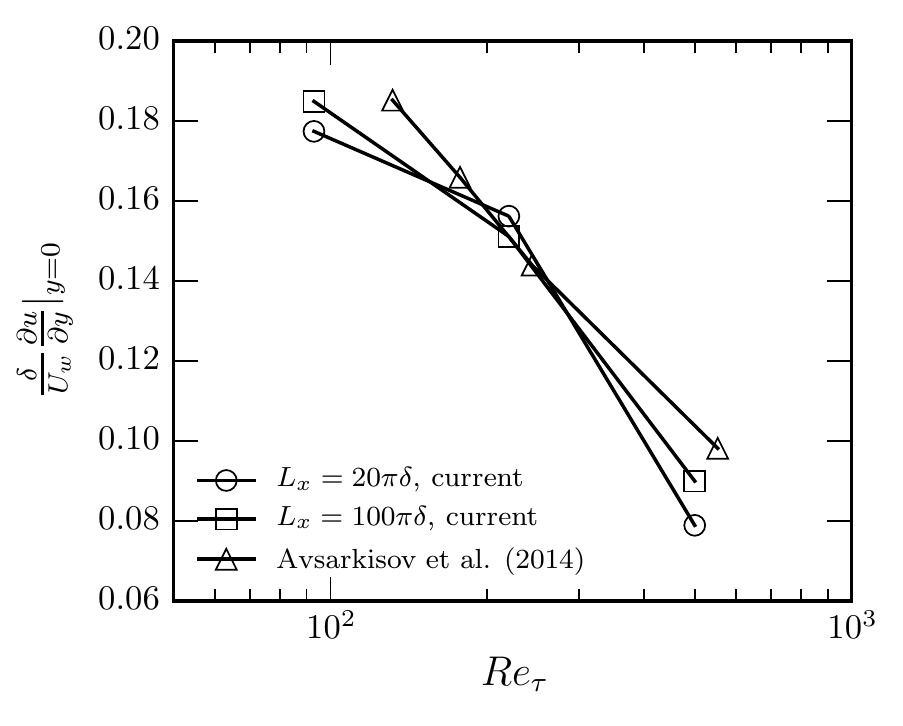}
  \caption{Mean velocity gradient at the center of channel}
  \label{fig:center_gradient}
\end{figure}

\begin{figure}
  \centering \includegraphics[width=\textwidth]{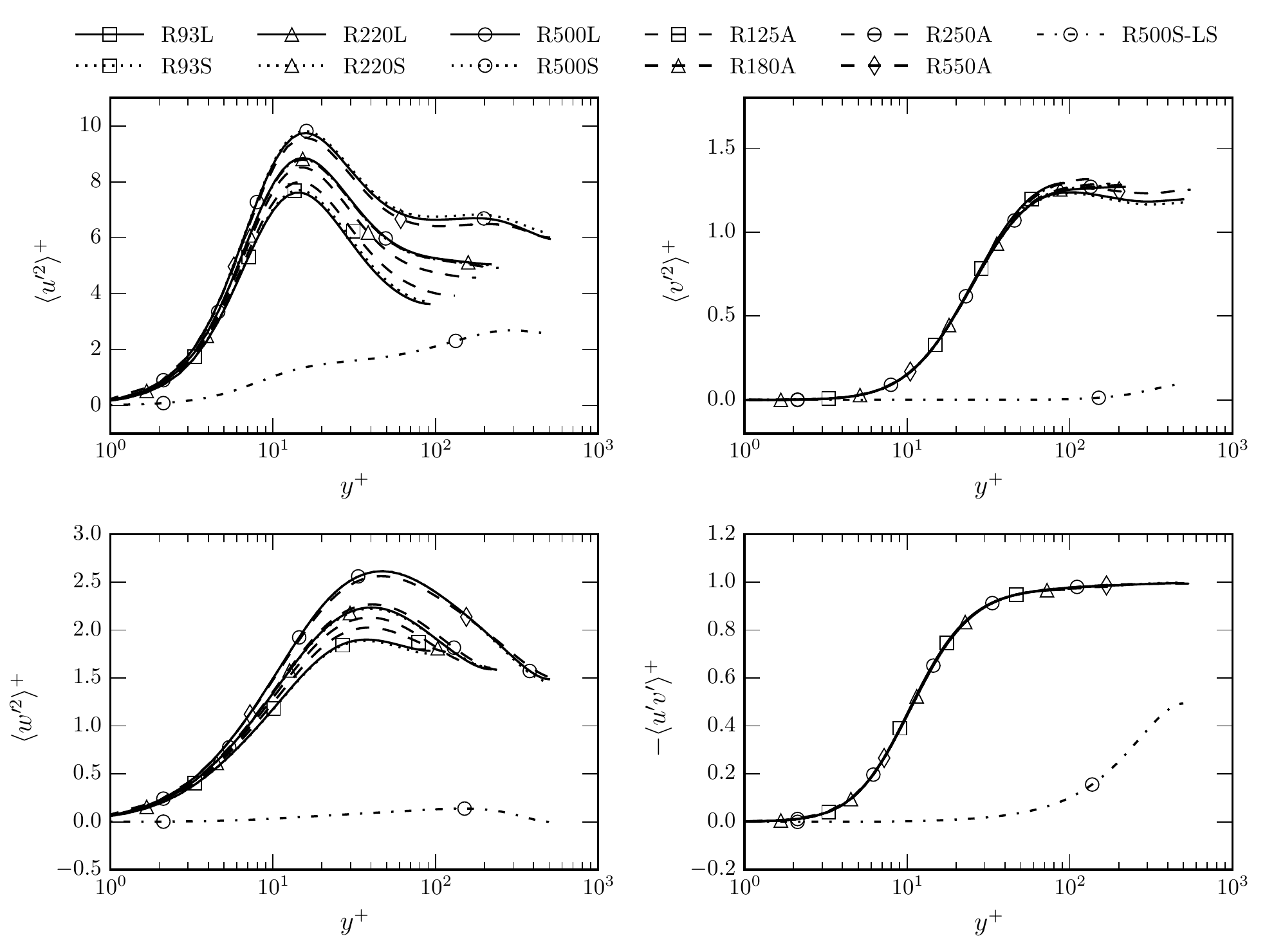} \caption{Profiles
  of the non-zero Reynolds stress components. Cases are defined in
  table~\ref{table:simulation_parameter}, and the R500S-LS curves are
  the contribution of large-scale structures to the Reynolds
  stress.}  \label{fig:uv}
\end{figure}

The derivative of the mean velocity in the center of the domain is
always zero in Poiseuille flows by symmetry. But in Couette flow the
mean velocity is anti-symmetric, and so the velocity gradient is not
necessarily zero at the center line, and therefore, neither is the
production of turbulent kinetic energy. The behavior of the mean
velocity gradient at the center as $Re$ increases is important to
understanding the production in the outer region, but as noted
by \citet{Avsarkisov:2014go} this remains
controversial \citep{Busse:1970gk,Lund:1980fs}.  Standard defect law
scaling with $u_\tau$ and $\delta$, and the common assumption that
$C_f=2u^2_\tau/U^2_w$ goes to zero as $Re\rightarrow\infty$ suggest
that the mean velocity gradient scaled by $U_w$ and $\delta$ should go
to zero at large Reynolds number. Indeed the data shown in
figure~\ref{fig:center_gradient} indicate that the gradient drops by
almost a factor of 3 over the factor of 5 increase in Reynolds number
investigated here. However, the Reynolds numbers are too low and the
range of Reynolds numbers too small to suggest the high Reynolds
number asymptotic behavior of the centerline derivative.  It is also
clear that there are modest differences in this derivative between the
small and large domain cases, as large as 14\% in the R500 case.
Further, there are differences in the center line gradient between the
current simulations and those of
\citet{Avsarkisov:2014go}, for unknown reasons. But this could be
due to large uncertainties in the statistics at the center of channel.

The non-zero Reynolds stress components appear in
Figure~\ref{fig:uv}. When plotted in wall units, many features are
similar to those in Poiseuille flow, especially near the wall, such as
the increase in the peak of $u'^2$ and to a lesser extent $w'^2$ with
Reynolds number. Away from the wall, particularly at the centerline,
$u'^2$ has a strong dependence on $Re$, which does not occur in
Poiseuille flow
\citep{Lee:2015er}. This may be related to the fact that the
production of $u'^2$ is zero in Poiseuille flow by symmetry, but not
Couette flow. Note that there is a mild secondary peak in $u'^2$ at
approximately $y^+=200$ in the R500 cases. Such an outer peak does not
occur in Poiseuille flow for $Re_\tau$ up to 5200 \citep{Lee:2015er},
but it is observed in both experiments at very high $Re$ in boundary
layers and pipes \citep{Hultmark:2012ce,Marusic:2010bb}. There is no
appreciable Reynolds number effect on $v'^2$ or $u'v'$ in these flows,
which is also different from Poiseuille flow. The effect of domain
size on the Reynolds stress components is negligible in most
cases. The one exception is $u'^2$, in which there is a modest
difference between the large and small domain cases away from the
wall. In general, the current Reynolds stress component data agree
with that from \citet{Avsarkisov:2014go}, with only minor
discrepancies, which are within the uncertainties described in
Appendix~\ref{sec:appendix}.

\begin{figure}
  \centering
  \begin{subfigure}{\linewidth}
    \includegraphics[width=\textwidth]{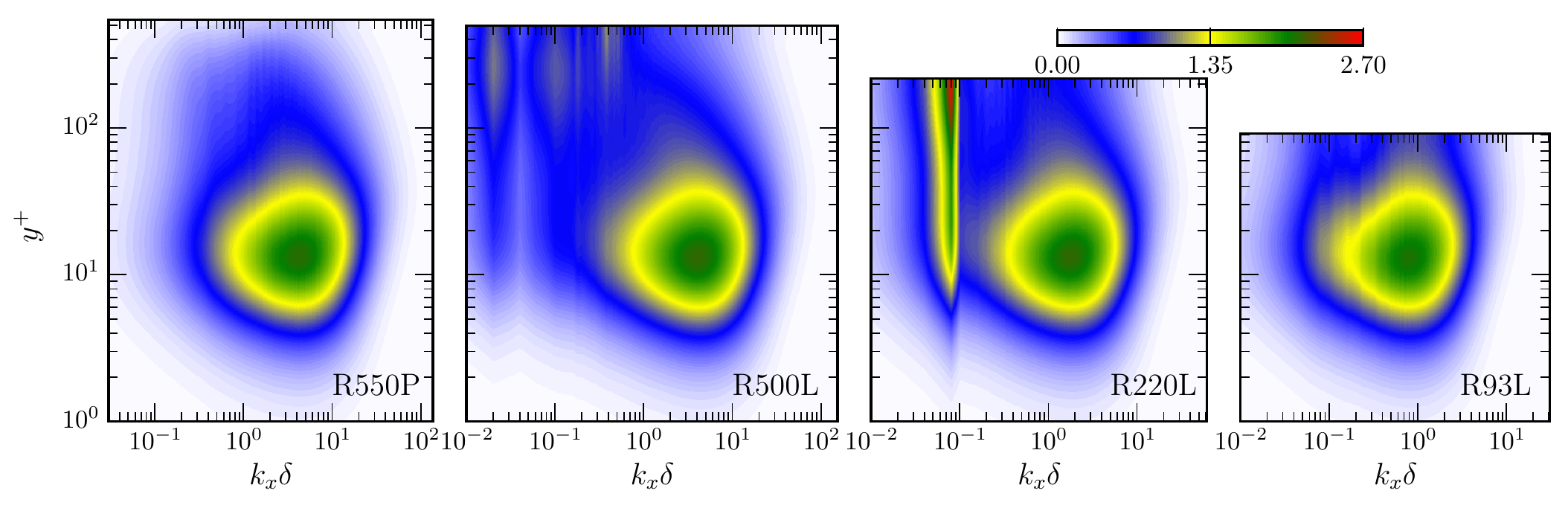}
    \caption{$k_x \delta E_{uu}/ u_\tau^2$}
    \label{fig:kxEuu}
  \end{subfigure} 
  \begin{subfigure}{\linewidth}
    \includegraphics[width=\textwidth]{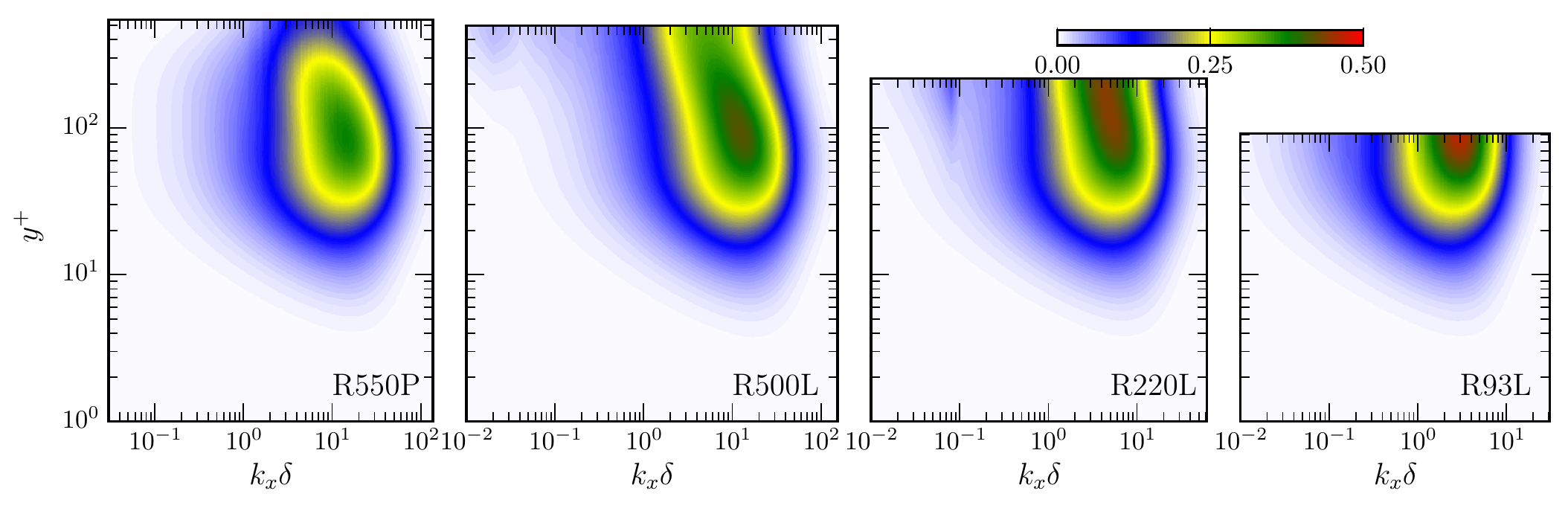}
    \caption{$k_x \delta E_{vv}/ u_\tau^2$}
    \label{fig:kxEvv}
  \end{subfigure} 
  \begin{subfigure}{\linewidth}
    \includegraphics[width=\textwidth]{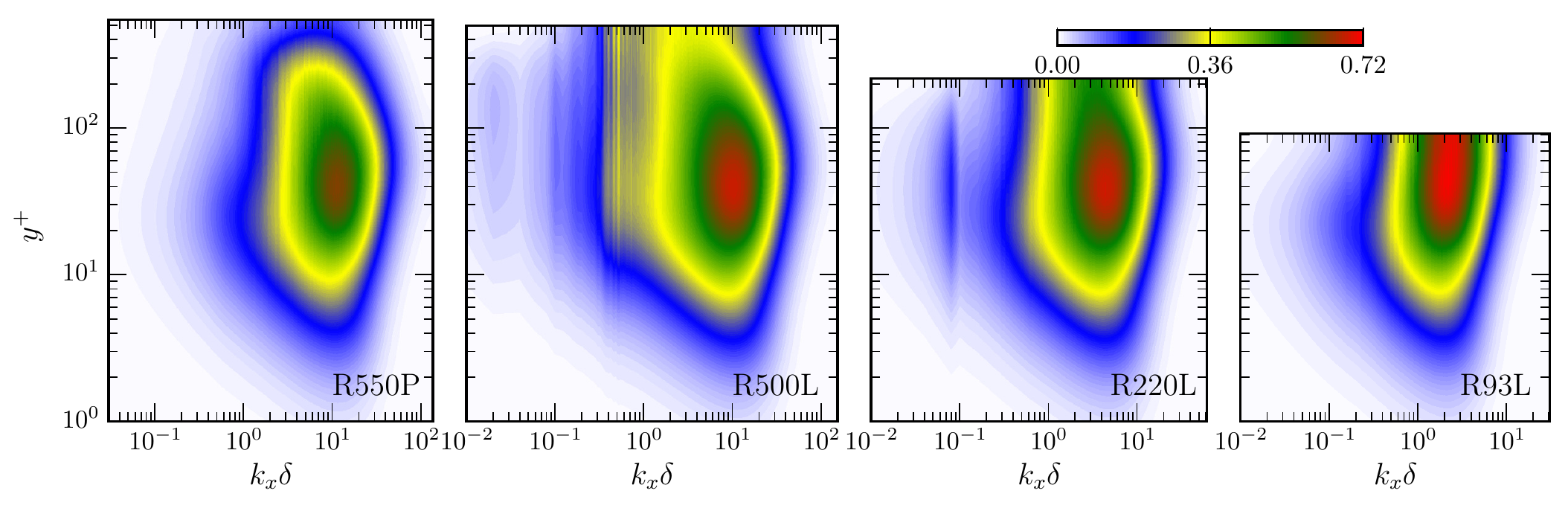}
    \caption{$k_x \delta E_{ww}/ u_\tau^2$}
    \label{fig:kxEww}
  \end{subfigure} 
  \begin{subfigure}{\linewidth}
    \includegraphics[width=\textwidth]{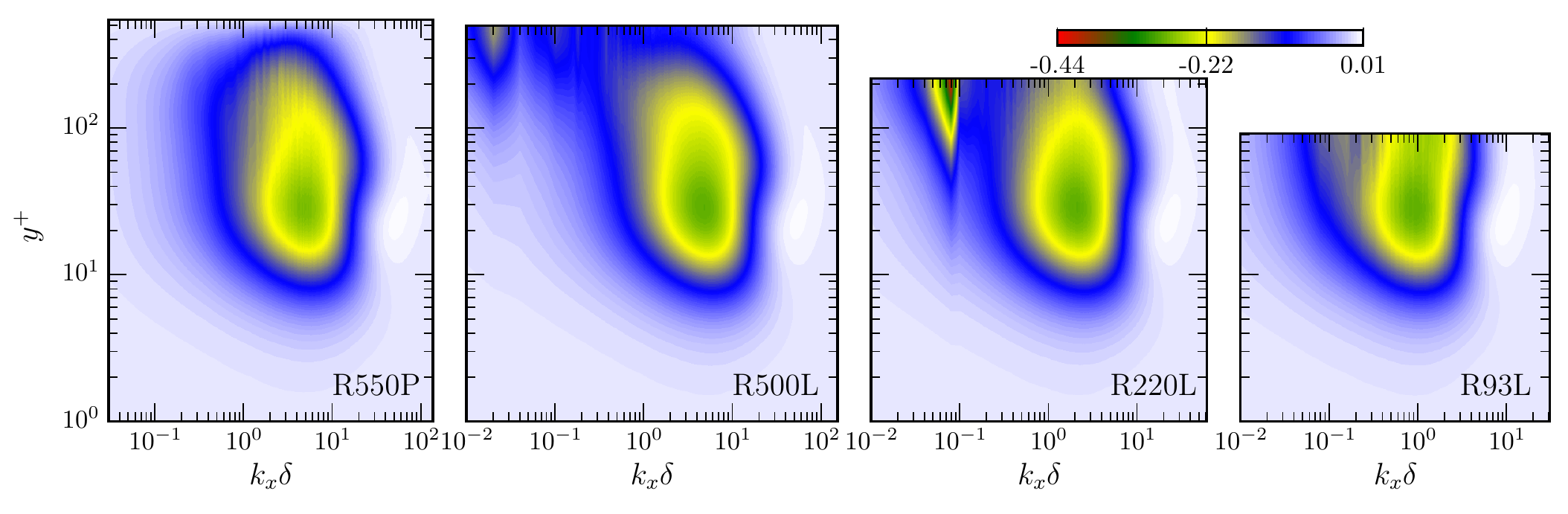}
    \caption{$k_x \delta E_{uv}/ u_\tau^2$}
    \label{fig:kxEuv}
  \end{subfigure} 
  \caption{(Colour online) Contours of the $k_x$-premultiplied
  streamwise one-dimensional spectra of $u_i' u_j'$ for the long
  domain Couette flow
  cases identified in table~\ref{table:simulation_parameter}, and the  
    planar Poiseuille flow of \citet{LozanoDuran:2014kr} at $Re_\tau =
  550$, labeled R550P.}
  \label{fig:kxE}
\end{figure}

\begin{figure}
  \centering
  \begin{subfigure}{\linewidth}
    \includegraphics[width=\textwidth]{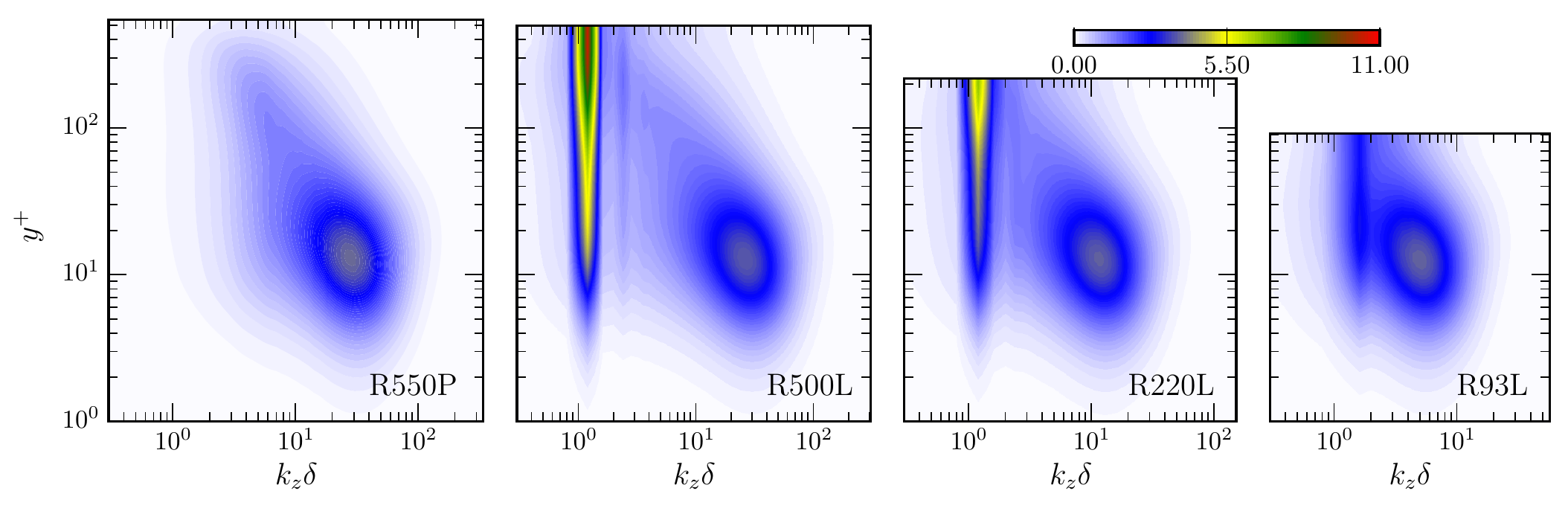}
    \caption{$k_z \delta E_{uu}/ u_\tau^2$}
    \label{fig:kzEuu}
  \end{subfigure} 
  \begin{subfigure}{\linewidth}
    \includegraphics[width=\textwidth]{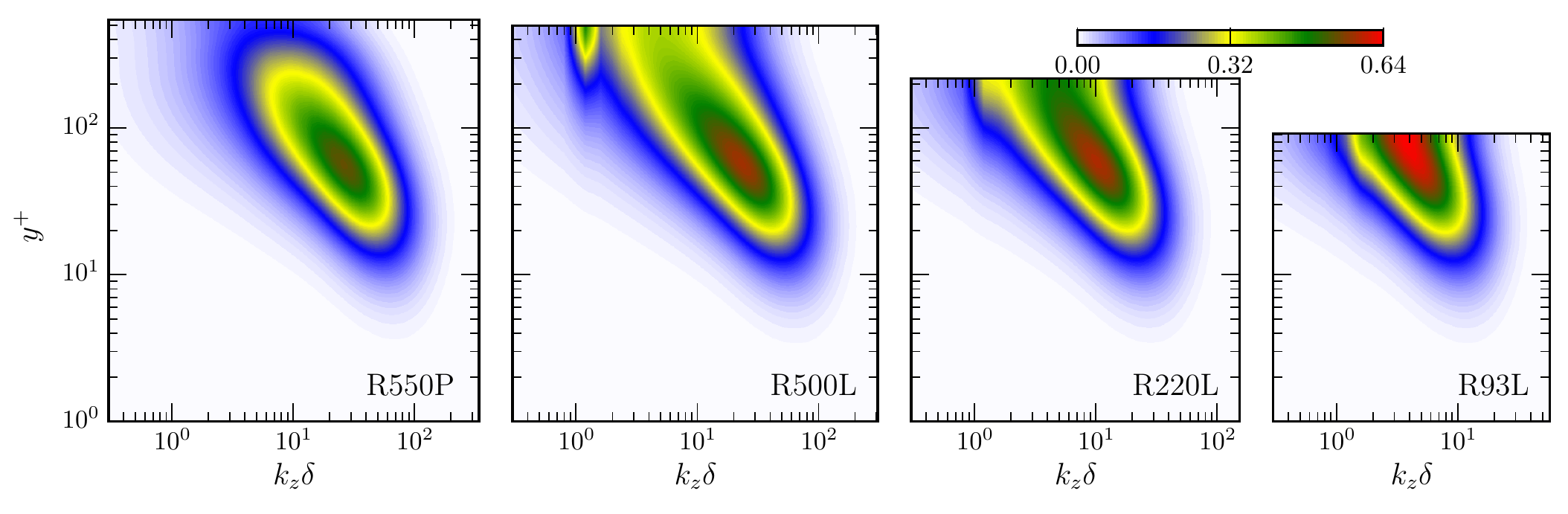}
    \caption{$k_z \delta E_{vv}/ u_\tau^2$}
    \label{fig:kzEvv}
  \end{subfigure} 
  \begin{subfigure}{\linewidth}
    \includegraphics[width=\textwidth]{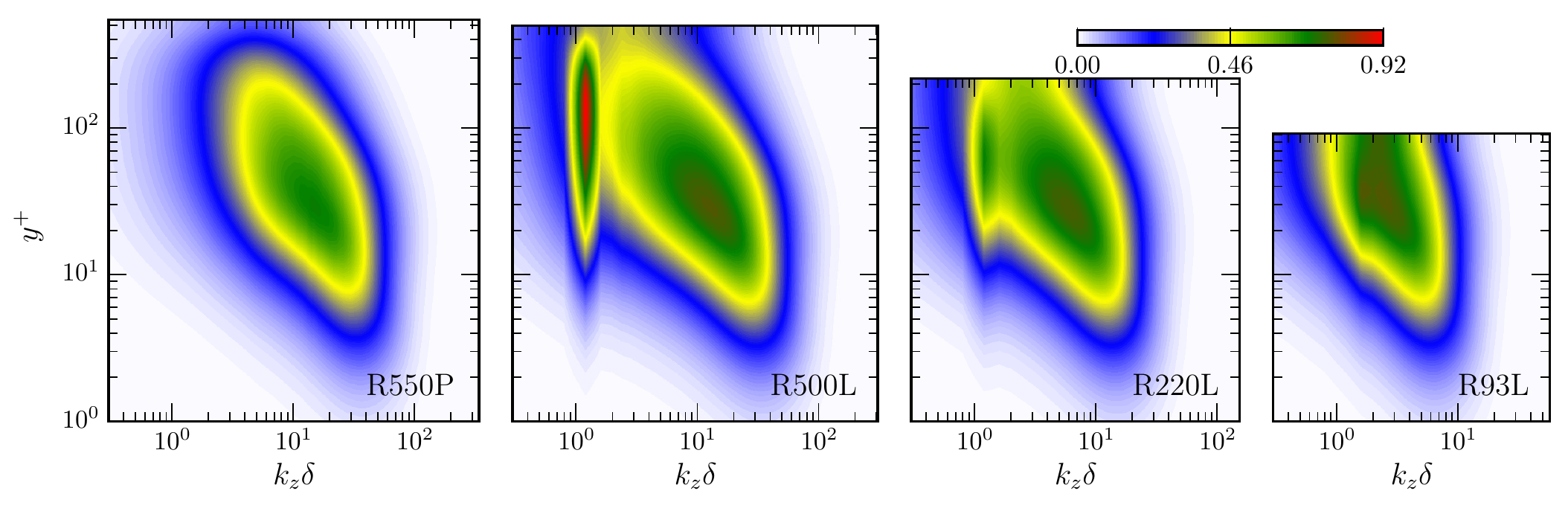}
    \caption{$k_z \delta E_{ww}/ u_\tau^2$}
    \label{fig:kzEww}
  \end{subfigure} 
  \begin{subfigure}{\linewidth}
    \includegraphics[width=\textwidth]{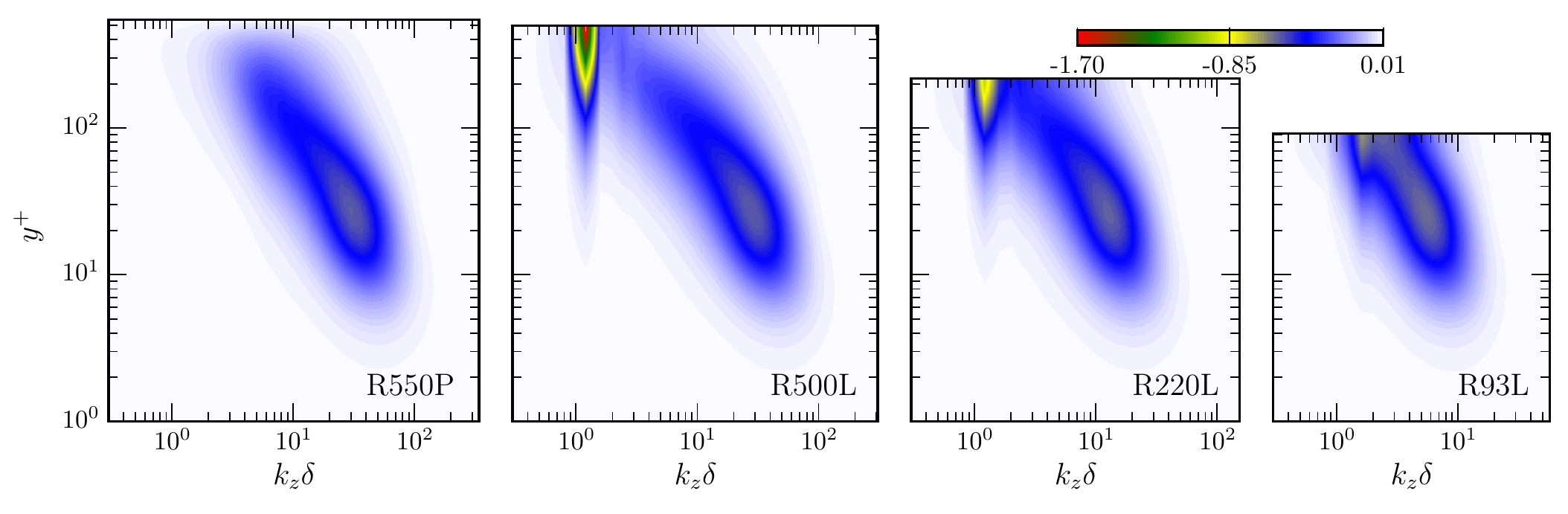}
    \caption{$k_z \delta E_{uv}/ u_\tau^2$}
    \label{fig:kzEuv}
  \end{subfigure} 
\caption{(Colour online) Contours of the $k_z$-premultiplied
  spanwise one-dimensional spectra of $u_i' u_j'$ for the long
  domain Couette flow
  cases identified in table~\ref{table:simulation_parameter}, and the  
    planar Poiseuille flow of \citet{LozanoDuran:2014kr} at $Re_\tau =
  550$, labeled R550P.}
  \label{fig:kzE}
\end{figure}

\begin{figure}
  \centering
  \begin{subfigure}{\textwidth}
    \includegraphics[width=\textwidth]{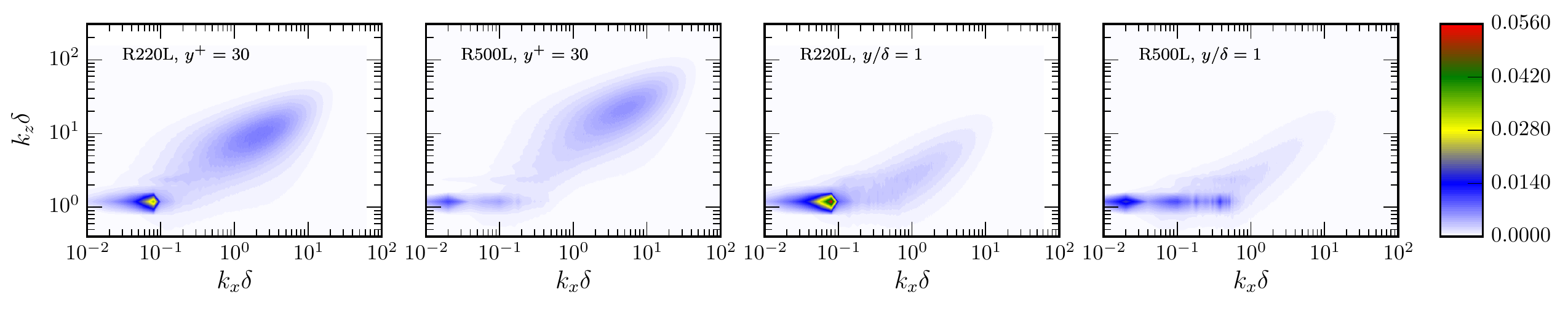}
    \caption{$k_x k_z \delta ^2 E_{uu}/ u_\tau^2$}
  \end{subfigure} 
  \begin{subfigure}{\textwidth}
    \includegraphics[width=\textwidth]{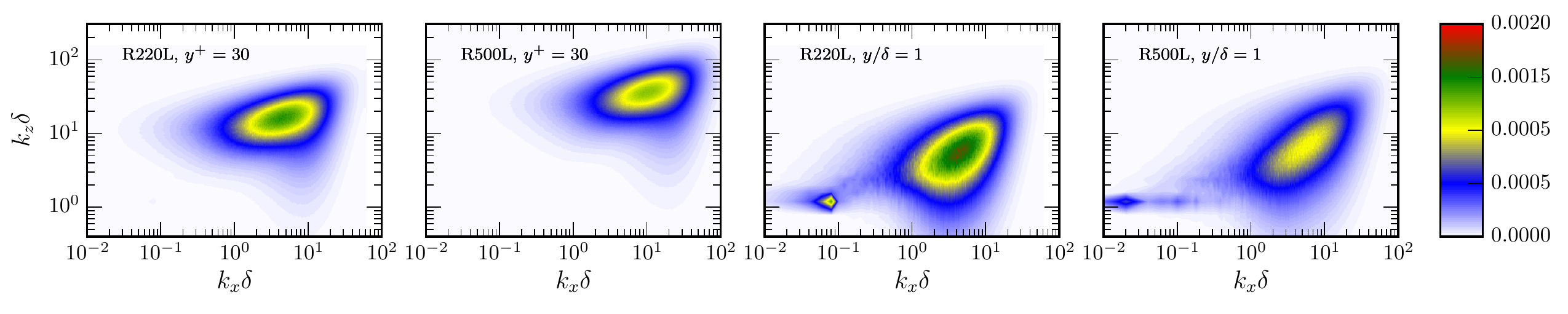}
    \caption{$k_x k_z \delta ^2 E_{vv}/ u_\tau^2$}
    \label{fig:E_vv_2d}
  \end{subfigure} 
  \begin{subfigure}{\textwidth}
    \includegraphics[width=\textwidth]{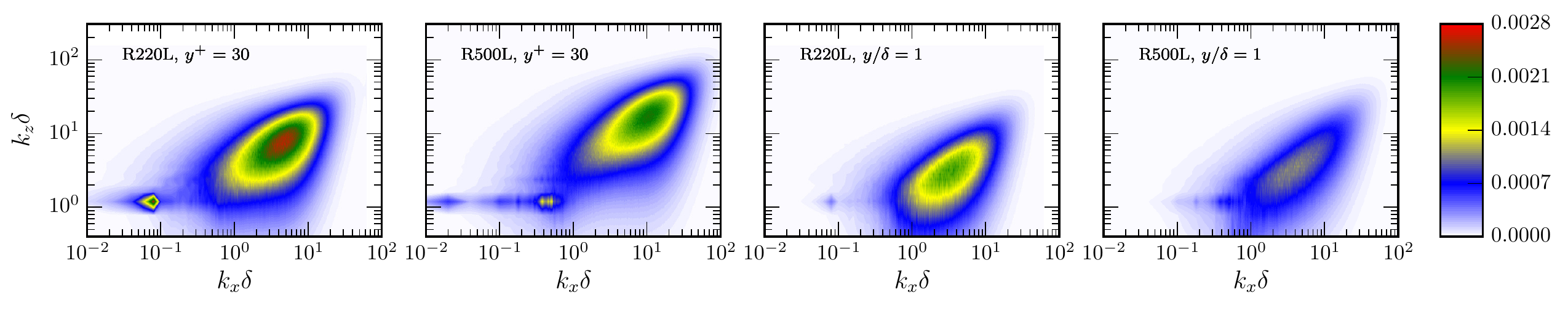}
    \caption{$k_x k_z \delta ^2 E_{ww}/ u_\tau^2$}
    \label{fig:E_ww_2d}
  \end{subfigure} 
  \begin{subfigure}{\textwidth}
    \includegraphics[width=\textwidth]{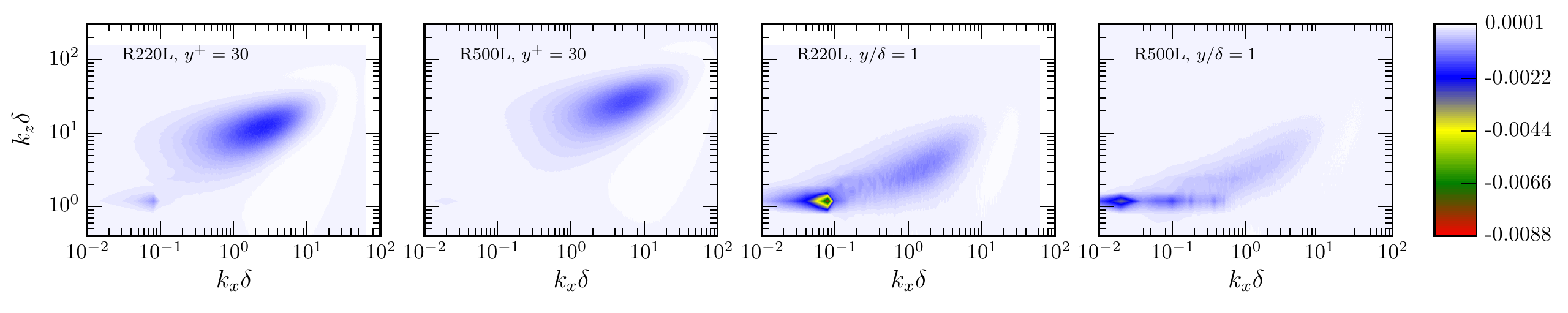}
    \caption{$k_x k_z \delta ^2 E_{uv}/ u_\tau^2$}
    \label{fig:E_uv_2d}
  \end{subfigure} 
  \caption{(Colour online) Contours of the $k_x$- and
  $k_z$-premultiplied two-dimensional spectra of $u'_i u'_j$ for the R220L and R500L cases identified in table 1}
  \label{fig:E_2d}
\end{figure}

While the effects of the computational domain size on the one-point
statistics is modest, the presence of large-scale turbulent eddies in
the larger domain is expected to have a large effect on the spectra,
which are shown as a function of $y$ in figures~\ref{fig:kxE} and
\ref{fig:kzE}. 
The so-called premultiplied spectrum $kE(k)$ is plotted because the
wavenumber axis is logarithmic and integrating $kE(k)$ over $d\log(k)$
yields the energy. Spectra of the non-zero components of the Reynolds
stress tensor, including the cross-spectra of $uv$ are shown for each
of the large domain cases. In addition, for comparison, we include the
spectra from the large-domain Poiseuille flow at $Re_\tau=550$
conducted by \citet{LozanoDuran:2014kr} (labeled R550P) with $L_x =
60 \pi\delta$ and $L_z = 6\pi\delta$. A striking feature of these
spectra is that for small enough scales (say $k_x^+ > 0.002$ and
$k_z^+> 0.02$) and close enough to the wall (say $y^+ < 60$), the
spectra of each Reynolds stress component are the same regardless of
Reynolds number and including the Poiseuille flow. This is due to the
approximate universality of small-scale near-wall turbulence.

Another feature of the Couette flow spectra is the sharp peak in the
spanwise ($k_z$) spectrum at around $k_z\delta\approx 1.2$. In
$E_{uu}$ and $E_{ww}$, this peak spans from the center line to around
$y^+\approx 5$, while in $E_{vv}$ and $E_{uv}$ it occurs only further
from the wall. This peak does not appear in the Poiseuille flow
(R550P). Note that peaks of $E_{uu}$ at low wavenumber in the outer
flow are observed in boundary layers and channels at higher $Re$ (say
$Re_\tau$ = 5000 or higher) in experiments and
DNS \citep{Hutchins:2007ty,Lee:2015er}. In addition, DNS data of
channel flow exhibits an outer peak of $E_{uv}$ in flows at higher
$Re$ \citep{Lee:2015er}. The outer peaks of $E_{uu}$ and $E_{uv}$ in
boundary layers and channels do not extend from the outer flow to near
the wall, and they are much broader in wavenumber. They seem to be a
different feature.  So the spectral peak observed here appears to be a
consequence of the large-scale motions specific to Couette flow.  The
spanwise scale of these structures is consistent with previous
observations by
others. \citep{Tsukahara:2006bg,Avsarkisov:2014go,Pirozzoli:2014bm}

In the streamwise Couette flow spectra, there is a corresponding peak
in $E_{uu}$ and $E_{uv}$ at $k_x\delta\approx 0.08$ at $Re_\tau=220$
and it is broadly consistent with the experiments of
\citet{Kitoh:2008gj,Kitoh:2005tn}. In the $Re_\tau=500$ case, the
$E_{uv}$ spectrum has a peak near the centerline at $k_x\delta=0.02$
and there are multiple weaker low wavenumber peaks in $E_{uu}$, with
the lowest also occurring at $k_x\delta\approx 0.02$. The energy in
the large-scale modes is thus distributed across a range of
wavenumbers. Note that $k_x\delta=0.02$ is the lowest non-zero
wavenumber in $x$, corresponding to a wavelength equal to the domain
size. At $Re_\tau=500$, both $E_{vv}$ and $E_{ww}$ also exhibit low
wavenumber peaks, though they are so weak they are difficult to detect
in the figure. These weak spectral peaks occur at $k_x\delta\approx
0.02$ and $0.1$. The streamwise extent of the structures in Couette
flow, as measured by the wavelength of these spectral peaks is quite
large, up to $310\delta$ at $Re_\tau=500$. And, the extent appears to
increase with Reynolds number, as it is only about $78\delta$ at
$Re_\tau=220$. Because the low wavenumber peak at $Re_\tau=500$ occurs
at the lowest represented wavenumber, the streamwise structure of the
these large-scale motions is impacted by the finite domain size, so
our assessment of their natural streamwise length-scale cannot be
definitive.

\begin{figure}
  \centering
    \includegraphics[width=\textwidth]{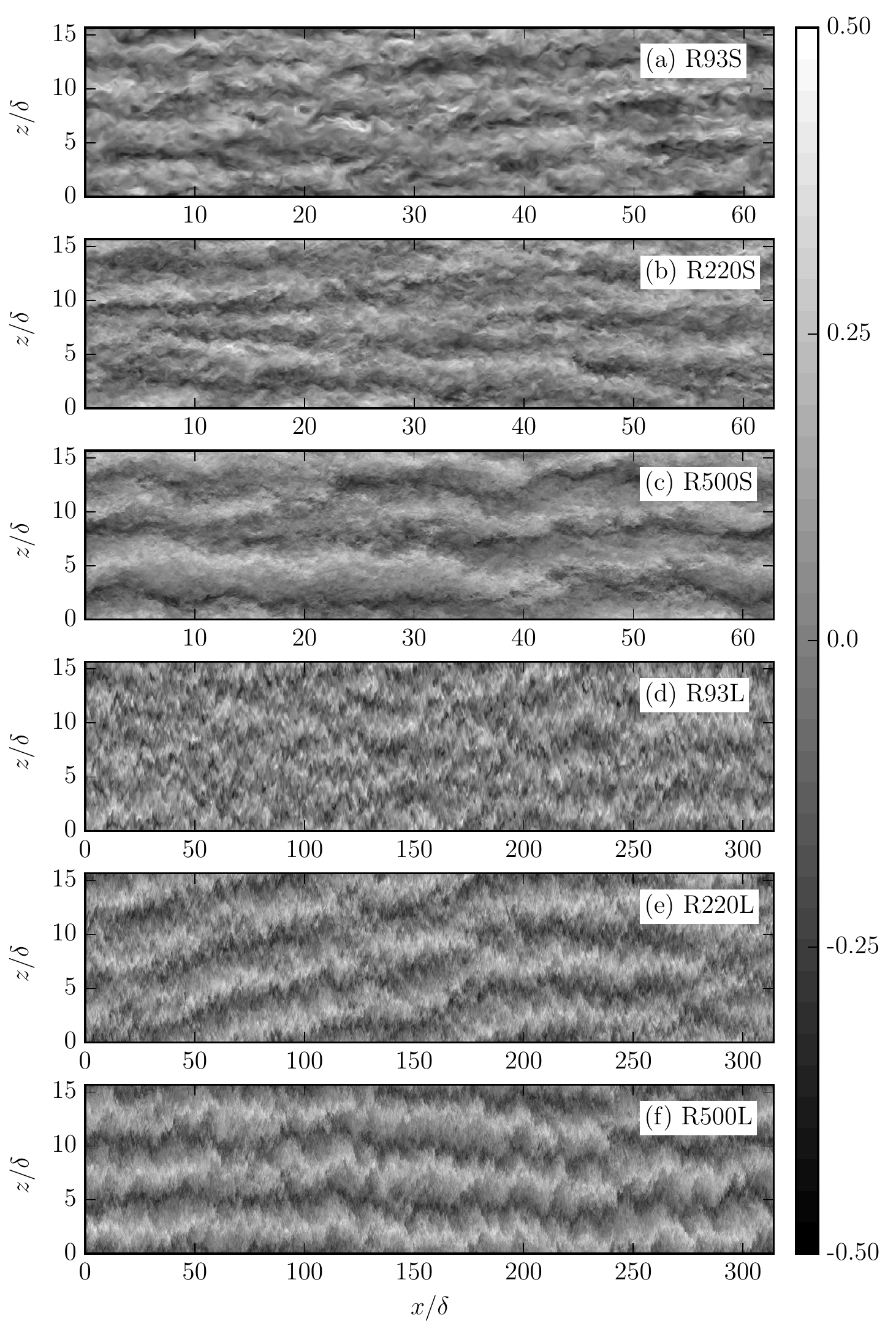}
  \caption{Instantaneous streamwise velocity in a wall-parallel
    plane in the center of the domain}
  \label{fig:U_field}
\end{figure}

Note that in the spanwise spectra, the large scale spectral peaks we
have been discussing have their maximum magnitude at the centerline,
except for $E_{ww}$, where it is very weak at the
centerline. Furthermore, these peaks extend much closer to the wall in
$E_{uu}$ and $E_{ww}$, than for $E_{vv}$. These features are
consistent with large-scale motions dominated by large-scale
streamwise vortices that extend from one wall to the other, as
proposed by \citet{Komminaho:1996fu,Papavassiliou:1997cy,Tsukahara:2006bg,Avsarkisov:2014go}.
Similar spectral indicators of large-scale streamwise vortices were
observed in transitional Couette flow \citep{Tuckerman:2011er}.

The two-dimensional energy spectra of the non-zero Reynolds
stress components are shown in figure~\ref{fig:E_2d}. Shown are two
$k_x$--$k_z$ planes with $y^+ = 30$ and $y/\delta = 1$.  As expected,
the dominant small-scale mass of the spectra shifts to larger
$k_x\delta$ and $k_z\delta$ (smaller scale) with increase Reynolds number.
In addition, the two-dimensional spectra have peaks at $k_z\delta \approx 1.2$ for
both the R220L and R500L cases. While the peak in the R220L case is
stronger and localized in streamwise scale at $k_x\delta \approx
0.08$, the peak in the R500L case is strongest at $k_x \delta \approx
0.02$, with energy distributed along a line at $k_z\delta \approx
1.2$, to $k_x\delta$ as low as 0.6. Note that the spectral peaks in
$E_{uu}$ and $E_{ww}$ near the wall are at the same wavenumbers as at
the center line. These observations are consistent with the
observations of the one-dimensional spectra.

The streamwise velocities of an instantaneous field in the
wall-parallel plane at the center of the domain are shown in
figure~\ref{fig:U_field}. Note that the scale is compressed by a
factor of 5 in the $x$ direction for the long domain cases. In all
cases, there are strong streaks in the opposite directions.  In the
R93S/L cases (figure~\ref{fig:U_field}a,d), the strong streaks are
observed but the length of streaky motion is somewhat too short to
find patterns.

In the R220L case (figure~\ref{fig:U_field}e), there are low and high
speed streaks, which are inclined from the streamwise
direction. Further, since the domain is periodic, these tilted streaks
have a helical topology, as do, presumably, the large vortical
structures underlying them. These tilted streaks are responsible for
the presence of the strong low-wavenumber peak in the R220L streamwise
spectrum (figure~\ref{fig:kzE}). In a shorter domain
(figure~\ref{fig:U_field}b), a larger tilt angle would be required to
produce a helical topology. Apparently, such strongly tilted
structures are not preferred, so in the smaller domain, the periodic
boundary conditions effectively ``lock'' the structures into an
untilted configuration. Note that rotating the simulation domain may
facilitate study of such tilted structures using simulations with
limited domain sizes \citep{Barkley:2005ko,Duguet:2010dv}.

In the higher Reynolds number R500L case (figure~\ref{fig:U_field}f),
there is no dominant inclination to the streaks, or helical
topology. Instead, they are predominantly streamwise, with large scale
meanders and distortions that are too large to be seen in the
short-domain case (figure~\ref{fig:U_field}c). Clearly, it is the lack
of a dominant inclination and the presence of the meanders that is
responsible for the more complex low-wavenumber streamwise spectrum
$E_{uu}$, with multiple weak peaks at the centerline.

\begin{figure}
  \centering
    \includegraphics[width=0.9\textwidth]{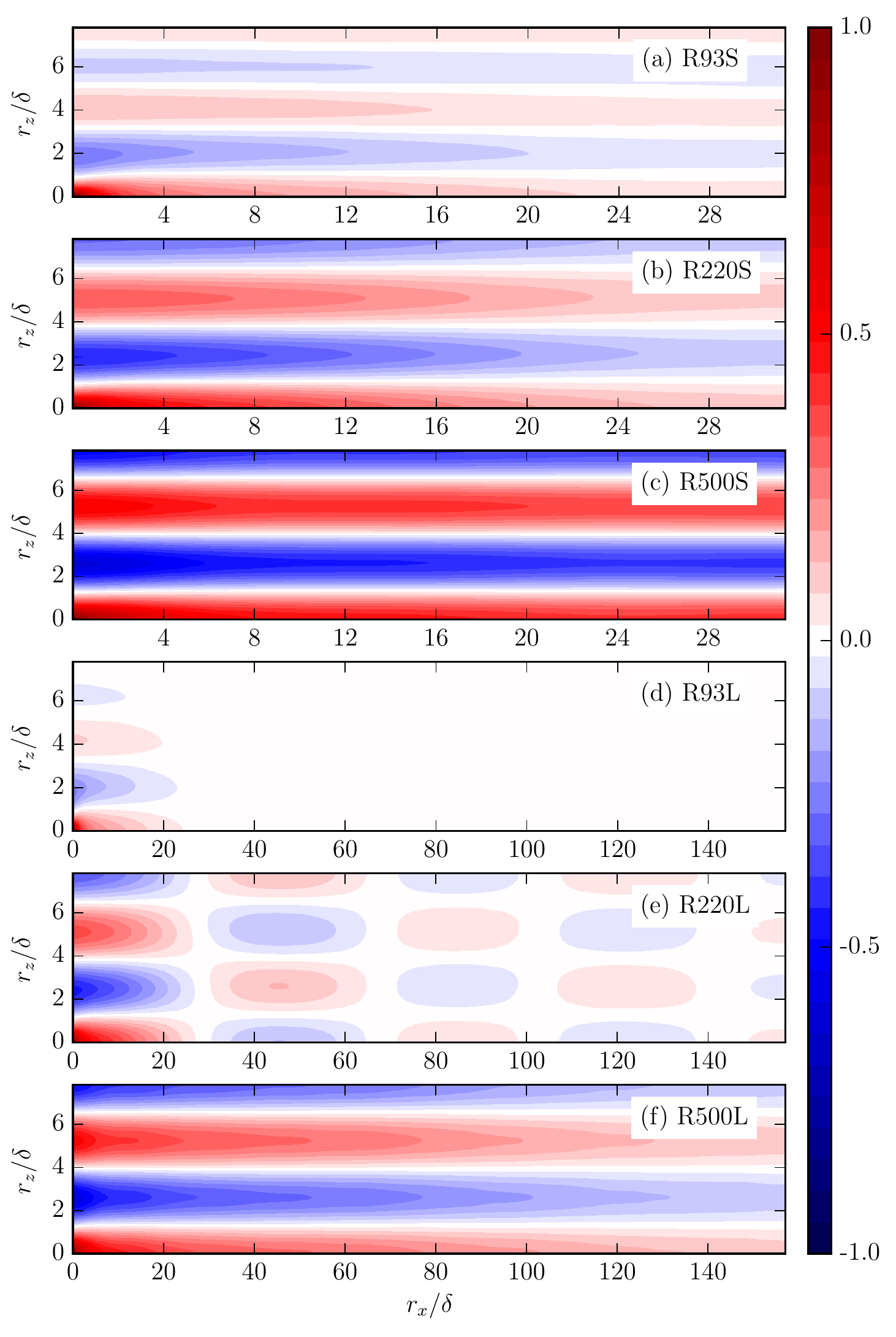}
  \caption{(Colour online) Two-point correlations of streamwise velocities in a wall-parallel
    plane in the center of the domain}
  \label{fig:2pt_corr}
\end{figure}

To quantitatively examine the nature of the large-scale structures
indicated in figure~\ref{fig:U_field}, the two-point
auto-correlation of the streamwise velocity,
$R_{uu}(r_x,y,r_z)=\langle u'(x,y,z)
u'(x+r_x,y,z+r_z)\rangle/\langle u'^2 \rangle$, for $y=0$ (the
centerline) is shown in figure~\ref{fig:2pt_corr} where $r_x$ and
$r_z$ are the separations in the streamwise and spanwise
directions, respectively. In general, the strength of the
correlation at large separations increases with $Re$ indicating
the increased coherence of the underlying structures. At the
highest Reynolds number ($Re_\tau = 500$), the underlying
structure is highly coherent in the spanwise direction as
indicated by the fact that the magnitude of the correlation
coefficient with $r_x=0$ and $r_z=5\pi\delta/2$ (half the spanwise
domain size) is about 0.5 in both the R500S and R500L cases.  In
all cases, there is a regular alternation in sign of the
correlation in the spanwise direction, with a wavelength of
$5\pi\delta/3$ in the R220 and R500 cases and $5\pi\delta/4$ in
the R093 case, which are one third or one fourth of the spanwise
domain size respectively. This is necessarily consistent with the
presence of the strong spectral peaks at $k_z\delta=1.2$ or 1.6 in
the spanwise spectra shown in figure~\ref{fig:kzE}. This suggests
that the spanwise scale of the underlying large-scale structures
is a function of Reynolds number. But, the wavelengths that can be
supported in the computational domain are highly constrained, so
it is not possible to determine the natural spanwise size of the
large-scale structures, or how it depends on Reynolds
number. However, given the nature of the underlying structures
(streamwise vortices filling the domain in the wall-normal
direction, see below), it is not expected that the spanwise
wavelength could become many times larger than observed here.

For all three Reynolds numbers, the short domain cases have
correlation coefficients that do not change sign over the entire
streamwise domain, as shown in
figures~\ref{fig:2pt_corr}a,b,c. However, at the lowest Reynolds
number the correlation falls almost to zero at maximum streamwise
separation (0.05 at $r_z=0$). In contrast, for the R500S case, the the
correlation falls only to 0.4 at $r_x=L_x/2$ and $r_z=0$. This is
consistent with the instantaneous streamwise velocities shown in
figure~\ref{fig:U_field}a,b,c, which have little streamwise coherence
for R93S, but significant coherence for R500S. In the long domain
cases, the correlation decays in the streamwise direction somewhat
faster than in the short domain cases. For example, in R93L, the
correlation with $r_z=0$ falls to less than 0.05 by $r_x=10\pi$, and,
in R500L, the correlation at $r_x=10\pi$ is about 0.3, rather than 0.4
in R500S. Apparently, the periodic boundary conditions in the short
domain help reinforce the structures and thus make them somewhat more
streamwise coherent.

In the R220L case, the correlations oscillate in the streamwise
direction as they do in the spanwise direction, in this case, with a
wavelength of $25\pi\delta$, which is 15 times the spanwise
wavelength.  The reason for this streamwise oscillation is the
inclination of the low and high-speed streaks shown in
figure~\ref{fig:U_field}e. A correlation pattern like that in
figure~\ref{fig:2pt_corr}e arises from such inclined structures when
inclinations of either positive or negative slope are equally
likely. The average slope of the inclination is just the ratio of the
spanwise to streamwise wavelengths or 1/15, which is approximately
$4^\circ$ from alignment in the streamwise direction.  As with the
spanwise wavelength, the finite domain size and periodic boundary
conditions in the streamwise direction constrain the possible
streamwise wavelengths, and thus the possible inclination slopes. In
this case, with a spanwise wavelength of $\frac53\pi\delta$, the
possible values of the slope are $\frac53n\pi\delta/L_x$. In the large
domain this is $n/60$ and in the small domain $n/12$. The inclination
slope arising in the R220L case is thus smaller than the smallest
non-zero slope possible in the R220S case. This may be why the
inclination is zero in R220S.

The question remains as to why there is an inclination to the
structures in the R220L case, but none is apparent in the correlations
for R93L or R500L. As discussed above, the coherence of the structures
underlying the correlations are strongly dependent on the Reynolds
number, and the R93L case appears to be too low in Reynolds number to
have sufficient streamwise coherence to exhibit inclined structures in
the correlations. For R500L, it may be that the preferred inclination
angle is also strongly dependent on $Re$ and at this Reynolds number
is too small to be represented in the long spatial domain (i.e. slope
less than $1/60$). Or, it could be that this tendency to generate
inclined structures is an intermediate Reynolds number phenomenon
only. In either case, it would seem that very long structures that are
significantly inclined from the streamwise direction are unlikely for
higher Reynolds numbers than those studied here.

\begin{figure}
  \centering
  \begin{subfigure}{\linewidth}
    \includegraphics[width=\textwidth]{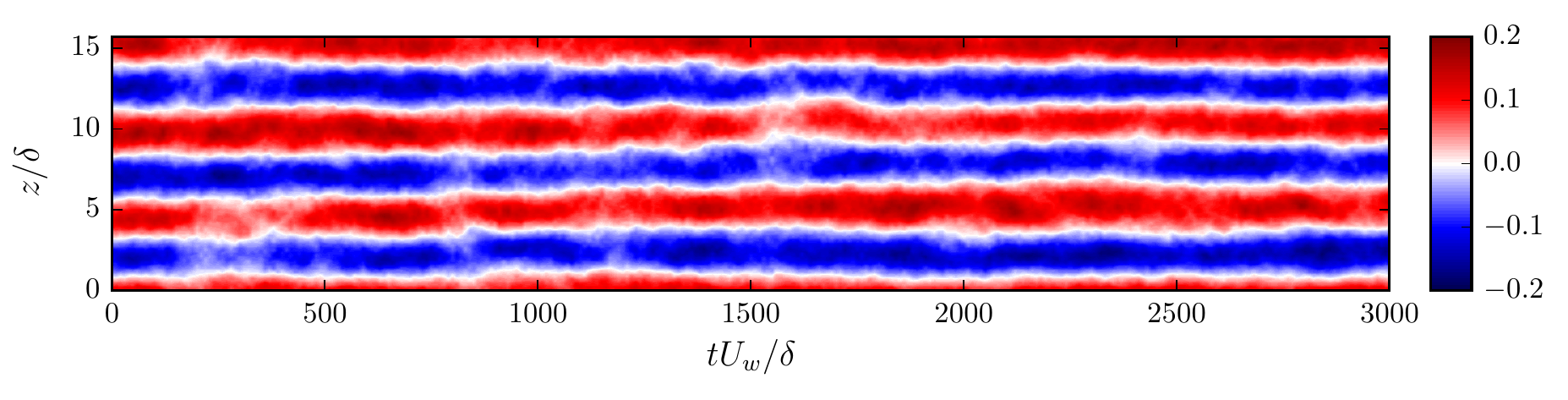}
    \caption{Streamwise velocity at the centerline averaged in the streamwise direction.}
    \label{fig:R500S_avg_u_t_vs_z}
  \end{subfigure} 
  \begin{subfigure}{\linewidth}
    \includegraphics[width=\textwidth]{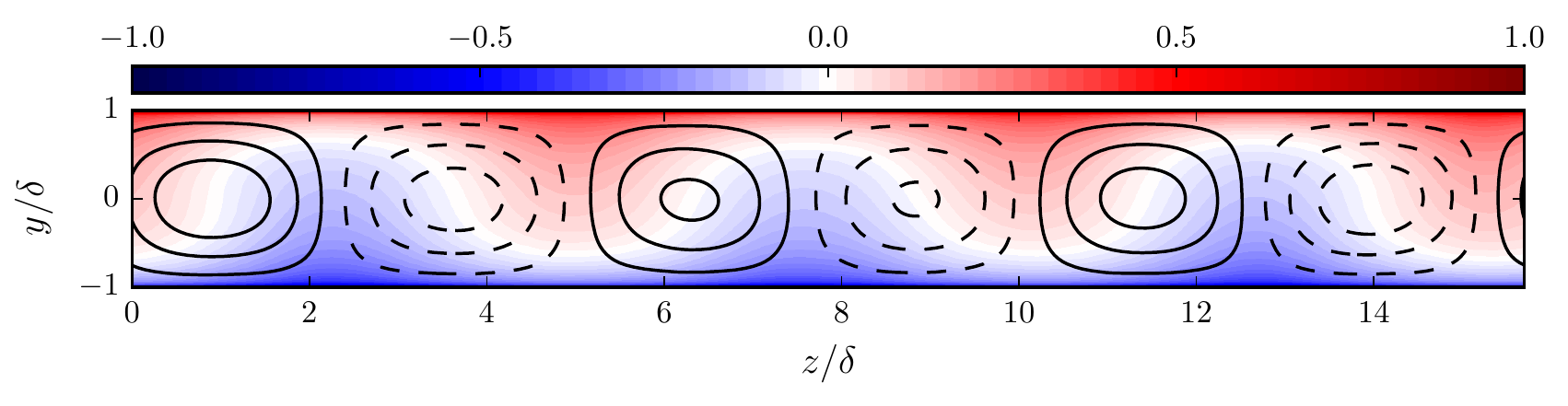}
    \caption{Velocity averaged in time and the streamwise
  direction. Shown are the streamwise velocity (colour) and the stream function; ------ (positive) -- -- -- (negative)}
    \label{fig:R500S_avg_u_y_vs_z}
  \end{subfigure} 
  \caption{(Colour online) Averaged velocity fields; R500S}
  \label{fig:R500S_avg}
\end{figure}

To investigate the structures underlying the correlations in
figure~\ref{fig:2pt_corr}, we take advantage of the fact that the
R500S case has streamwise coherence that extends through the entire
streamwise computational domain, and average the velocity in the
streamwise direction in this case. The coherence of this streamwise
averaged velocity in time is characterized by the contours of the
averaged streamwise velocity as a function of $z$ and time, as shown
in figure~\ref{fig:R500S_avg}a. Note that these velocities do appear
to be coherent in time, and exhibit no substantial spanwise drift over
the time simulated ($3000\delta/U_w$). This suggests that an average
in the streamwise direction and time would yield a good representation
of the structures responsible for the observed correlations. The
results are shown in figure~\ref{fig:R500S_avg}b, where the streamwise
velocity and contours of the stream function are plotted. As expected,
and as observed previously by
others \citep{Papavassiliou:1997cy,Pirozzoli:2014bm,Avsarkisov:2014go},
the structures consist of counter-rotating streamwise vortices that
fill the domain between the moving walls. Embedded as they are in the
background shear of the Couette flow, they produce large fluctuations
in the streamwise velocity. The maximum wall-normal and spanwise
velocities associated with these vortices are approximately $0.025U_w$
and $0.03U_w$, respectively, while at the center of the domain, the
streamwise velocity varies between $\pm 0.13U_w$.  These structures
make significant contributions to $\langle u'^2\rangle$ and $\langle
u'v'\rangle$, and as much as half of the total Reynolds shear stress
at the center of the domain (see R500S-LS profiles in
figure~\ref{fig:uv}). The vortices are also clearly
responsible for the low wavenumber features of the spectra in
figures~\ref{fig:kxE}--\ref{fig:E_2d} and discussed above.
Since the streamwise and temporal averaging has
reduced the apparent strength of the structures by averaging out
meandering and variability, this estimate of contributions to Reynolds
stress are likely understated.

While the large-scale coherent vortices described above are a dominant
feature of the turbulent Couette flow, there is no evidence of
analogous coherent vortices, perhaps spanning from the wall to the
center line, in turbulent Poiseuille flow (see case R550P in
figures~\ref{fig:kxE}--\ref{fig:E_2d}, for example). The reason for this difference is not
clear. Some insight might be gained by simulation and analysis of
model versions of turbulent Couette and Poiseuille flow that eliminate
the complexity of the wall by introducing mean shear through
volumetric forcing, as in \citet{Waleffe1997} and \citet{Chantry:2016cr}, since it seems
unlikely that the near-wall layer plays an important role in the
formation, or not, of these coherent vortices.

\section{Conclusions}
\label{sec:conclusion}
The spectra, flow visualizations and averaged structures presented
here indicate the presence of large-scale structures in Couette flow,
consisting of predominantly streamwise counter-rotating vortices that
extend from near the wall on one side to near the wall on the other,
as has been observed by
others \citep{Komminaho:1996fu,Papavassiliou:1997cy,Tsukahara:2006bg,Avsarkisov:2014go}. However,
there is a qualitative difference in the streamwise structure of these
vortices in the large domain DNS at $Re_\tau=220$ and 500. At the
lower Reynolds number, the structures are inclined from the streamwise
direction by an average of about $4^\circ$, while this is not the case
at the higher Reynolds number. The reason for this difference is
unknown. The large-scale vortices at $Re_\tau=500$ also have much
greater streamwise coherence, with the coherence length exceeding the
domain size ($100\pi\delta$). At $Re_\tau=93$ and 220, the coherence
length is much smaller (of order $20\delta$ and $100\delta$
respectively).  Thus, it is clear that the streamwise coherence length
of the large-scale vortices grows rapidly with Reynolds number.

Current results show that the Reynolds stress components have strong
$Re$ dependencies. It is also the case that the small-scale motions,
as characterized by the spectra, are quantitatively consistent among
cases with different Reynolds numbers. Therefore, one can conclude
that the $Re$ dependencies in the Reynolds stress components are the
result of the increasing role of large-scale motions, which is
consistent with the observation that the average large scale
structures make a significant contribution to the streamwise velocity
variance and the Reynolds shear stress; as much as half of the latter.
The large contribution to the mean momentum flux at the centerline by
these structures that span the domain from wall to wall may be the
reason for the sharp reduction of the mean velocity gradient near the
center at $Re_\tau=500$, as observed
by \citet{Avsarkisov:2014go,Orlandi:2015bl} and shown in
figure~\ref{fig:profile}.

In the present work, we compare the data from simulations in domains
with $L_x=100\pi\delta$ and $20\pi\delta$. One-point statistics shows
no significant difference between the cases with different domain
sizes, despite the qualitative differences in spectra and large-scale
structure in these cases. This may be a consequence of the relatively
low Reynolds numbers, with the observed differences in large-scale
features of the flows in different domain sizes having an increasing
impact on the statistics as Reynolds number increases.  Studying this
further will require simulations at higher Reynolds numbers and in
longer domains, making it out of reach for DNS for now. However, large
eddy simulation (LES) may be a good alternative, since the small
scales are remarkably consistent across the flows studied here.

The data presented in this paper are available on line
at \url{http://turbulence.ices.utexas.edu}.

\begin{figure}
  \centering
  \begin{subfigure}{\textwidth}
  \includegraphics[width=\textwidth]{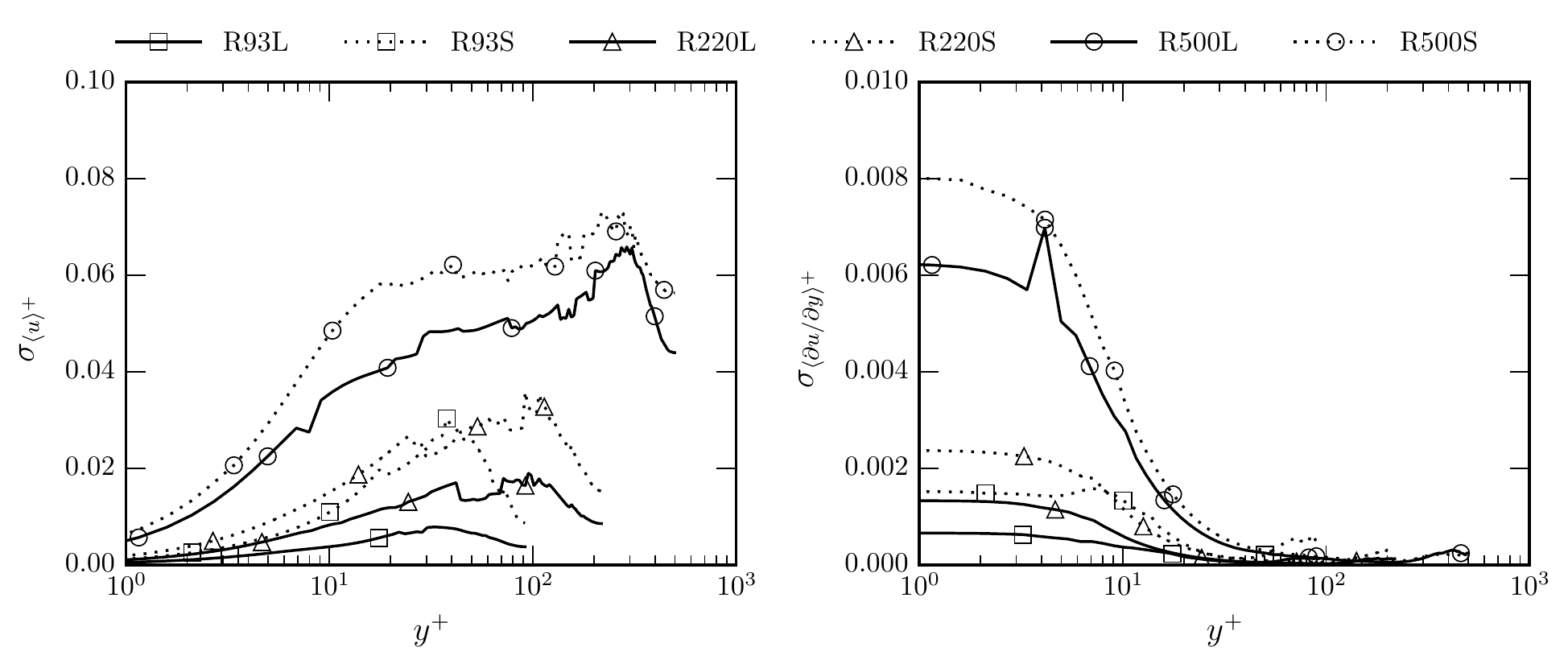}
  \caption{Mean velocity and its gradient}
  \label{fig:app_U_sigma}
  \end{subfigure}
  \begin{subfigure}{\textwidth}
    \centering
    \includegraphics[width=\textwidth]{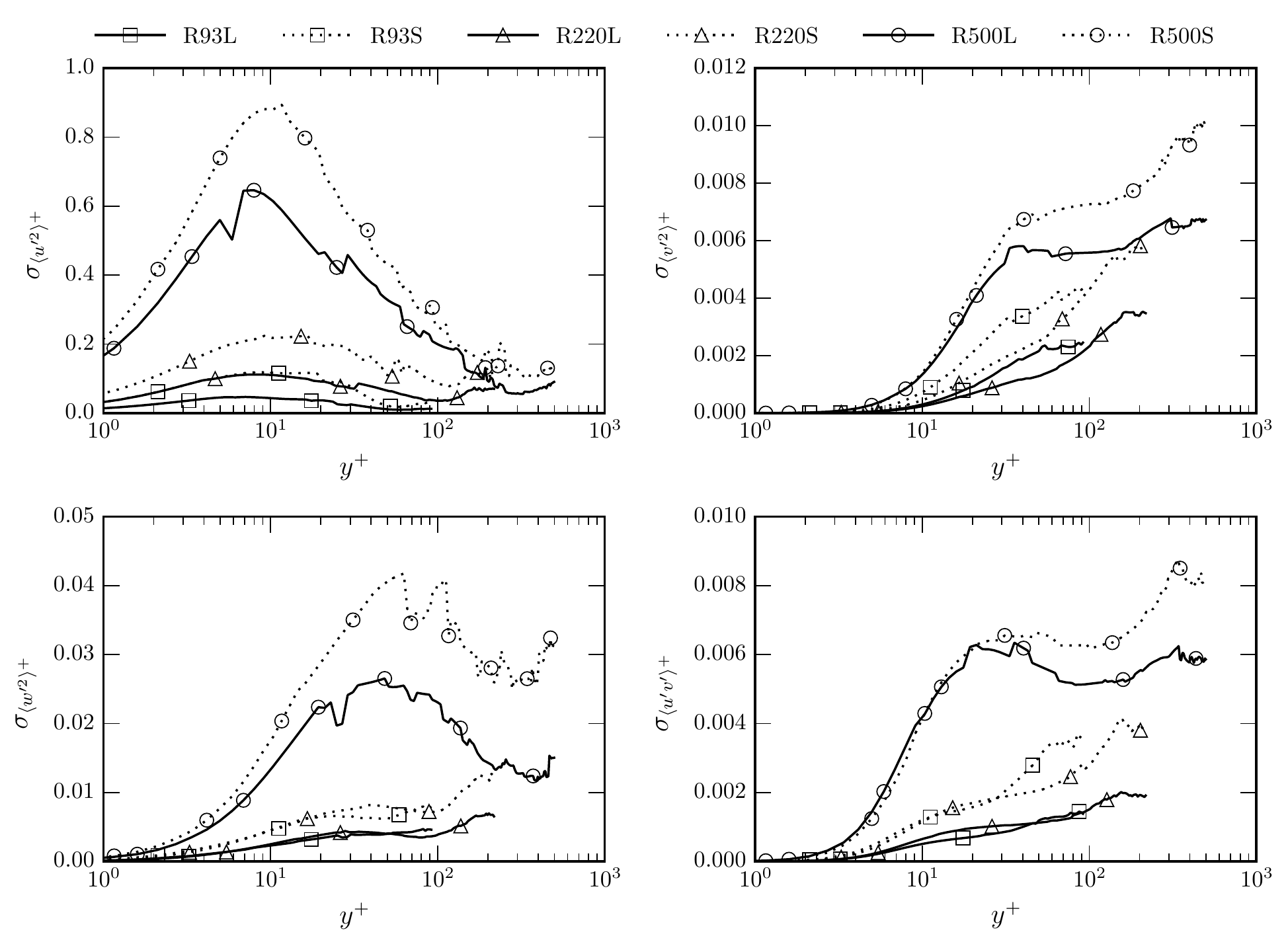}
    \caption{Non-zero Reynolds stress components}
    \label{fig:app_uv_sigma}
  \end{subfigure}
  \caption{Standard deviations of estimated statistical quantities due
    to finite sample size; Cases are defined in table~\ref{table:simulation_parameter}.}
  \label{fig:app_sigma}
\end{figure}

\section*{Acknowledgments}
We are very grateful to Dr. Adri\'an Lozano-Dur\'an and Prof. Javier
Jim\'enez who kindly shared their DNS data.  This research used
resources of the Argonne Leadership Computing Facility, which is a DOE
Office of Science User Facility supported under Contract
DE-AC02-06CH11357 and Texas Advanced Computing Center (TACC) at The University of Texas at Austin.

\appendix
\section{Estimated uncertainties of one-point statistics}
\label{sec:appendix}
The estimated uncertainties in the mean velocity, velocity
gradients and Reynolds stress components reported here, due to
aveaging over finite samples, are shown inf
figure~\ref{fig:app_sigma}. These uncertainties were estimated using
the technique described by \citet{Oliver:2014dh}.

\bibliographystyle{../sty/jfm}

\end{document}